\documentclass[letterpaper, 10 pt,conference]{ieeeconf}  % Comment this line out if you need a4paper
\let\labelindent\relax
\usepackage{amssymb}
\usepackage{amsmath,mathrsfs,empheq}
\usepackage{color}
\usepackage{graphicx}
\usepackage{enumitem}
\usepackage{mathtools}
\usepackage{lineno,hyperref}
\usepackage{cleveref}
\usepackage{bm}
\usepackage{tikz}
\usepackage{algorithm}
\usepackage{algorithmicx}
\usepackage{algpseudocode}
\modulolinenumbers[5]
\newtheorem{theorem}{Theorem}[section]
\newtheorem{remark}{Remark}[section]

\newtheorem{lemma}[theorem]{Lemma}

\newtheorem{problem}{Problem}[section]

\usepackage{multirow}
\usepackage{color}
\usepackage{cite}
\usepackage{comment}

\providecommand{\keywords}[1]{\textbf{\textit{Index terms---}}}

\IEEEoverridecommandlockouts                              % This command is only needed if 
\title{\LARGE \bf Truncated Gaussian Noise Estimation in State-Space Models}

\author{Rodrigo A. Gonz\'alez, Angel L. Cede\~no, Koen Tiels and Tom Oomen % <-this % stops a space
\thanks{This work was supported in part by the grant ANID-Fondecyt 3240181, by the ANID-Basal Project AFB240002 (AC3E), and by E2TECH, University of Santiago of Chile. R. A. Gonz\'alez, K. Tiels and T. Oomen are with the Mechanical Engineering Department of Eindhoven University of Technology, Eindhoven, The Netherlands. A. L. Cede\~no is with the University of Santiago of Chile (USACH), Faculty of Engineering, Electrical Engineering Department, Chile. T. Oomen is also with Delft Center for Systems and Control, Delft	University of Technology, The Netherlands.}%
}

\usepackage{eso-pic}
\AddToShipoutPictureBG*{%
	\AtPageUpperLeft{%
		\setlength\unitlength{1in}%
		\hspace*{\dimexpr0.5\paperwidth\relax}
		\makebox(0,-1.25)[c]{
			\begin{tabular}{c c}
				Rodrigo A. Gonz\'alez \emph{et al.},
				Truncated Gaussian Noise Estimation in State-Space Models, \\
				accepted for publication at IEEE Conference on Decision and Control (CDC) 2025, uploaded to ArXiv on July 25, 2025 \\
\end{tabular}}}}

\begin{document}

\maketitle

%%%%%%%%%%%%%%%%%%%%%%%%%%%%%%%%%%%%%%%%%%%%%%%%%%%%%%%%%%%%%%%%%%%%%%%%%%%%%%%%
\begin{abstract}
	Within Bayesian state estimation, considerable effort has been devoted to incorporating constraints into state estimation for process optimization, state monitoring, fault detection and control. Nonetheless, in the domain of state-space system identification, the prevalent practice entails constructing models under Gaussian noise assumptions, which can lead to inaccuracies when the noise follows bounded distributions. With the aim of generalizing the Gaussian noise assumption to potentially truncated densities, this paper introduces a method for estimating the noise parameters in a state-space model subject to truncated Gaussian noise. Our proposed data-driven approach is rooted in maximum likelihood principles combined with the Expectation-Maximization algorithm. The efficacy of the proposed approach is supported by a simulation example.
\end{abstract}
\begin{keywords}
Noise estimation; Truncated Gaussian distribution; EM Algorithm; Nonlinear System Identification.\end{keywords}

\section{Introduction}
The goal of system identification is to derive a mathematical model of a dynamical system using input-output data \cite{soderstrom1989}. A critical aspect of this process involves quantifying the uncertainty associated with the estimated model, as it directly impacts analysis, prediction, and robust control performance. These uncertainties may arise from structural simplifications in the model formulation, inaccuracies in parameter estimation, and measurement noise present during system identification experiments \cite{Maybeck1982,ljung1999}. Consequently, accurately modeling noise and additive uncertainties remains a fundamental challenge, significantly influencing both model precision and robustness. Traditionally, two dominant paradigms have been employed for uncertainty modeling: (1) the probabilistic approach, which characterizes noise using Gaussian probability density functions (PDFs), and (2) the set-membership approach, a distribution-free method that assumes noise to be deterministically bounded \cite{Milanese2013}. While the probabilistic framework provides a statistical interpretation of uncertainties, it often fails to account for hard constraints on noise magnitudes. Conversely, the set-membership approach incorporates prior knowledge in the form of amplitude bounds but does not directly consider statistical insights that could enhance model accuracy. 

Incorporating signal constraints, such as bounded disturbances, into the estimation process enhances the accuracy and reliability of system state estimates \cite{alamo2005guaranteed}, thereby supporting better decision-making in areas such as control, system identification, signal processing, and navigation. In system identification specifically, incorporating signal constraints strengthens the physical interpretability of solutions and facilitates the integration of prior knowledge about unknown parameters, leading to improved generalization performance.

Constraints on system noise have been incorporated into system identification through the set-membership approach \cite{fogel1982value}, which assumes amplitude-bounded noise. This framework has led to the development of various distribution-free model estimators \cite{hjalmarson1994unifying,hjalmarsson1998optimally,akccay2006membership}. However, existing efforts have primarily focused on system estimators, with limited attention given to parametric characterizations of bounded noise in state-space representations. Conversely, state-space modeling has predominantly relied on Gaussian assumptions for state and output noise \cite{gibson2005robust,yuz2011,wills2013,cedeno2024identification}. Since the Gaussian PDF has unbounded support, these approaches inherently disregard noise amplitude constraints, which may compromise the accuracy of uncertainty characterization. Other methods such as \cite{kost2022measurement} explore correlation methods for noise identification, with focus on noise moment estimation in linear systems.

A promising intermediate solution bridges the set-membership and Gaussian paradigms by employing truncated Gaussian PDFs. This approach enables more realistic noise modeling while retaining the mathematical tractability and interpretability of Gaussian distributions. Applications span diverse fields: autonomous robotics, where sensor noise is bounded but non-uniform \cite{Durrant-Whyte2006}; industrial control, where disturbances have operational limits \cite{Bittanti1994}; and biomedical systems such as glucose monitoring, where sensor errors must be constrained \cite{hovorka2004nonlinear}. Truncated statistics have also been studied in regression and classification problems \cite{ilyas2020theoretical}, probability distribution fitting \cite{lee2012algorithms}, and have been used as prior distributions for non-parametric system identification~\cite{Zheng2021}.

To bridge the gap between the probabilistic and set-membership approaches in noise estimation, this paper introduces a hybrid modeling strategy that integrates the strengths of both methods by employing truncated Gaussian distributions to parametrize the noise statistics in state-space models. This formulation allows for probabilistic reasoning while respecting known physical bounds on disturbances, and can be particularly important for systems where noise exhibits both a probabilistic structure and hard physical limits. 

In summary, the main contributions of this work are:
\begin{enumerate}[label=C\arabic*]
	\item
	\label{contributionC1}
        We propose an iterative procedure, based on the EM algorithm \cite{dempster1977maximum} to compute the noise parameters of a state-space model subject to truncated Gaussian noise. In particular, our approach enables the estimation of the mean and covariance of the Gaussian prior to truncation, as well as the upper and lower truncation limits. Our method requires particle smoothing \cite{doucet2000sequential} to match the sample and population moments of the truncated Gaussian noise at each EM iteration. 
	\item We show the effectiveness of the proposed approach through extensive Monte Carlo runs performed on a simulation example.
\end{enumerate}

The paper is structured as follows. In Section \ref{sec:problemformulation} we present the problem setup, and in Section \ref{sec:mlem} we revisit the maximum likelihood and EM iterations. The proposed identification method under truncated Gaussian noise PDFs is derived in Section \ref{sec:truncated}, while Section \ref{sec:implementation} discusses implementation aspects. A simulation example can be found in Section \ref{sec:simulations}, and Section \ref{sec:conclusions} provides concluding remarks.

\textit{Notation}: All vectors and matrices are written in bold, and all vectors are column vectors, unless transposed. %The identity matrix is denoted as $\mathbf{I}$, and its dimension can be deduced from context. 
The set of vectors $\{\boldsymbol{a}_i\}_{i=1}^N$ is denoted as $\boldsymbol{a}_{1:N}$. If $\boldsymbol{a}$ and $\boldsymbol{b}\in \mathbb{R}^n$ with elements $a_i$ and $b_i$ respectively, then the notation $\boldsymbol{a}\leq \boldsymbol{b}$ corresponds to the element-wise inequality $a_i\leq b_i$ for all $i\in \{1,\dots,n\}$. Depending on the context, we also refer to these elements as $\{\boldsymbol{a}\}_i$ and $\{\boldsymbol{b}\}_i$. %The operation $\textnormal{vec}\{\cdot\}$ denotes vectorization, and $\otimes$ denotes the Kronecker product. The boundary of the set $\mathcal{A}\subset \mathbb{R}^n$ is denoted $\partial \mathcal{A}$,
The support of the function $f(\cdot)$ is denoted as $\textnormal{supp}(f(\cdot))$, and $\mathbf{1}_{[\boldsymbol{a},\boldsymbol{b}]}(\mathbf{x})$ is the indicator function of the hyperrectangle in $\mathbb{R}^n$ with two vertices $\boldsymbol{a}$ and $\boldsymbol{b}$, i.e., $\mathbf{1}_{[\boldsymbol{a},\boldsymbol{b}]}(\mathbf{x})=1$ if $\boldsymbol{a}\leq \mathbf{x} \leq \boldsymbol{b}$, and zero otherwise.

\section{System setup and problem formulation}
\label{sec:problemformulation}
Consider the following discrete-time state-space system
\begin{subequations}
\label{dtsystem}
\begin{align}
    \label{dtsystem1}
    \mathbf{x}_{t+1} &= \mathbf{f}_t(\mathbf{x}_t,\mathbf{u}_t)+\mathbf{w}_t \\
    \label{dtsystem2}
    \mathbf{y}_t &= \mathbf{g}_t(\mathbf{x}_t,\mathbf{u}_t)+\mathbf{v}_t,
\end{align}    
\end{subequations}
where $\mathbf{x}_t$, $\mathbf{u}_t$, and $\mathbf{y}_t$ denote the state, input, and output, respectively, with dimensions $\mathbf{x}_t \in \mathbb{R}^n$, $\mathbf{u}_t \in \mathbb{R}^m$, and $\mathbf{y}_t \in \mathbb{R}^p$. The functions $\mathbf{f}_t(\cdot)$ and $\mathbf{g}_t(\cdot)$ are assumed known and map to the appropriate dimensions. These functions may be given by a white-box physical model \cite{ljung1994modeling}, or may also be available from a previous system identification step. For simplicity, we assume the initial state $\mathbf{x}_1$ is known, which is reasonable when the system starts from rest or when transient data are removed. Otherwise, it can be estimated from data using an extended approach beyond the scope of this paper.

\begin{figure}
	\centering{
		\includegraphics[width=0.3\textwidth]{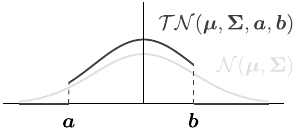}
\vspace{-0.3cm}
    \caption{Untruncated Gaussian (gray), and its truncated version (black), with lower and upper truncation bounds $\boldsymbol{a}$ and $\boldsymbol{b}$, respectively.}
\vspace{-0.4cm}
		\label{fig_truncated}}
\end{figure}

We may also write the system in \eqref{dtsystem} in the compact form
\begin{equation} \notag
\bm{\xi}_t = \mathbf{h}_t(\bm{\zeta}_t) + \bm{\eta}_t,
\end{equation}
with
\begin{equation}
    \bm{\xi}_t \hspace{-0.03cm}=\hspace{-0.03cm} \begin{bmatrix} \mathbf{x}_{t+1} \\ \mathbf{y}_{t} \end{bmatrix}, \hspace{0.1cm} \bm{\zeta}_t \hspace{-0.03cm}=\hspace{-0.03cm} \begin{bmatrix} \mathbf{x}_t \\ \mathbf{u}_{t} \end{bmatrix}, \hspace{0.1cm} \mathbf{h}_t(\cdot) \hspace{-0.03cm}=\hspace{-0.03cm} \begin{bmatrix} \mathbf{f}_t(\cdot) \\ \mathbf{g}_t(\cdot) \end{bmatrix}, \hspace{0.1cm} \bm{\eta}_t \hspace{-0.03cm}=\hspace{-0.03cm} \begin{bmatrix} \mathbf{w}_t \\ \mathbf{v}_{t} \end{bmatrix}\hspace{-0.03cm}. \hspace{-0.05cm} \notag
\end{equation}
The process and measurement noise, which are arranged in the vector $\bm{\eta}_t$, are assumed to be truncated normal distributed and uncorrelated with $\mathbf{x}_1$. The noise vector is described by 
\begin{equation}
    \bm{\eta}_t \sim \mathcal{TN}(\bm{\mu},\bm{\Sigma},\boldsymbol{a},\boldsymbol{b}),
\end{equation}
see Fig. \ref{fig_truncated} for a graphical interpretation. The probability density function (PDF) of $\bm{\eta}_t$ is given by
\begin{equation}
\label{pdftruncated}
    p(\bm{\eta}_t) =\displaystyle\frac{\exp\{-\frac{1}{2}(\bm{\eta}_t\hspace{-0.05cm}-\hspace{-0.05cm}\bm{\mu})^{\top}\bm{\Sigma}^{-1}(\bm{\eta}_t\hspace{-0.05cm}-\hspace{-0.05cm}\bm{\mu})\}}{\int_{\boldsymbol{a}-\bm{\mu}}^{\boldsymbol{b}-\bm{\mu}}\exp\{\hspace{-0.02cm}-\frac{1}{2}\mathbf{p}^{\top}\bm{\Sigma}^{-1}\mathbf{p}\}\textnormal{d}\mathbf{p}}\mathbf{1}_{[\boldsymbol{a},\boldsymbol{b}]}(\bm{\eta}_t).
\end{equation}
The quantities $\bm{\mu}$ and $\bm{\Sigma}$ correspond to the mean and covariance of the unconstrained normal distribution, and the lower and upper bounds of the truncated normal distribution are defined as $\boldsymbol{a}$ and $\boldsymbol{b}$, respectively. Note that the PDF describing $\bm{\eta}_t$ differs from that of a censored Gaussian, where the probabilities of the events ${\bm{\eta}_t = \boldsymbol{a}}$ and ${\bm{\eta}_t = \boldsymbol{b}}$ are strictly positive for finite-valued $\boldsymbol{a}$ and $\boldsymbol{b}$.

The problem of interest is stated next.
\begin{problem}
\label{problem1}
Given $N$ input-output samples $\{\mathbf{u}_t,\mathbf{y}_t\}_{t=1}^N$, and assuming that $\mathbf{f}_t$ and $\mathbf{g}_t$ in \eqref{dtsystem} are known or previously estimated, our goal is to obtain estimates for the parameters of the truncated normal noise distribution, including the truncation limits. The parameters of interest are arranged in a vector $\bm{\beta}$ of the form
\begin{equation}
\label{parameterbeta}
\bm{\beta} = [\bm{\mu}^\top, \textnormal{vec}\{\bm{\Sigma}\}^\top, \boldsymbol{a}^\top, \boldsymbol{b}^\top]^\top.
\end{equation}
\end{problem}

The next section introduces the statistical tools that are used for solving this problem.

\section{Maximum Likelihood and the EM algorithm}
\label{sec:mlem}
It is widely known that the maximum likelihood (ML) method provides consistent and asymptotically efficient estimators in many practical settings \cite{lehmann2006theory}. Such method seeks the parameter vector that maximizes the likelihood that the output data takes the observed values. More formally, the ML estimator is obtained as
\begin{equation}
    \label{likelihood}
    \hat{\bm{\beta}}_{\textnormal{ML}} = \underset{\bm{\beta}}{\arg \max}~\ell(\bm{\beta}),
\end{equation}
where $\ell(\bm{\beta}) = \log p(\mathbf{y}_{1:N}|\bm{\beta})$. This optimization problem is challenging to solve even in the linear system and unconstrained Gaussian case, since it involves a non-convex cost function that requires robust implementations to be adopted for obtaining the global maximum \cite{gibson2005robust}. One way to address this problem is to apply the Expectation-Maximization (EM) algorithm \cite{dempster1977maximum}, which provides a sequence of estimates $\hat{\bm{\beta}}^{(k)}$, $k=1,2,\dots$, that monotonically increase the likelihood, converging under general conditions to a local or global maximum of $\ell(\bm{\beta})$ if properly initialized~\cite{wu1983convergence}.

The EM algorithm requires defining latent variables, which, if known, could ease the optimization of the likelihood function. When studying state-space systems, it is natural to take the state sequence $\mathbf{x}_{1:N+1}$ as the latent variables \cite{yuz2011,gonzalez2023algorithm}. Thus, we consider the joint PDF
\begin{align}
    \mathcal{L}(\bm{\beta})&=p(\mathbf{x}_{1:N+1},\mathbf{y}_{1:N}|\bm{\beta}) \notag \\
    \label{mathcall}
    &=p(\mathbf{x}_1)\prod_{t=1}^{N} p(\mathbf{x}_{t+1},\mathbf{y}_t|\mathbf{x}_t,\bm{\beta}),
\end{align}
where the Markov property of the state-space formulation has been exploited. The EM algorithm is described in Algorithm \ref{algorithmem}. In practice, the EM algorithm terminates when a maximum number of iterations has been reached, or when the relative error of the estimates is below a predefined tolerance factor.
\begin{algorithm}
	\caption{EM algorithm for the computation of $\hat{\bm{\beta}}_{\textnormal{ML}}$}
	\begin{algorithmic}[1]
		\State Choose an initial estimate $\hat{\bm{\beta}}^{(1)}$
		\For{$k = 1, 2, \dots$}
		\State \textbf{E-step}: Compute the expectation
		\begin{equation}
			\label{qfunction2}
			\mathcal{Q}(\bm{\beta},\hat{\bm{\beta}}^{(k)}) = \mathbb{E}\left\{\log \mathcal{L}(\bm{\beta})\big| \mathbf{y}_{1:N},\hat{\bm{\beta}}^{(k)} \right\},
		\end{equation} 
            where $ \mathcal{L}(\bm{\beta})$ is given by \eqref{mathcall}.
		\State \textbf{M-step}: Solve the optimization problem
		\begin{equation}
			\label{mstep}
			\hat{\bm{\beta}}^{(k+1)} = \underset{\bm{\beta}}{\arg \max}~\mathcal{Q}(\bm{\beta},\hat{\bm{\beta}}^{(k)}).
		\end{equation}
		\EndFor
	\end{algorithmic}
	\label{algorithmem}
\end{algorithm}
\section{Parameter estimation of truncated Gaussian disturbances}
\label{sec:truncated}

In this section, we propose the algorithm to compute the maximum likelihood estimator for $\bm{\beta}$, the parameters associated with the truncated Gaussian noise distribution. To this end, we address the computation of the $\mathcal{Q}$ function that aims at solving Problem~\ref{problem1}, and we provide a characterization of its maximizer for each iteration of the EM algorithm.

For a known $\mathbf{x}_1$, the $\mathcal{Q}$ function is proportional to
\begin{equation}
\label{qfunction}
    \mathcal{Q}(\bm{\beta},\hat{\bm{\beta}}^{(k)}) \hspace{-0.03cm}\propto \hspace{-0.04cm} \sum_{t=1}^N \hspace{-0.03cm} \mathbb{E} \hspace{-0.01cm}\left\{ \log\hspace{-0.02cm} p(\mathbf{x}_{t+1}\hspace{-0.02cm},\hspace{-0.02cm}\mathbf{y}_t|\mathbf{x}_t,\hspace{-0.02cm}\bm{\beta})\big|\mathbf{y}_{1\hspace{-0.01cm}:\hspace{-0.01cm}N}\hspace{-0.02cm},\hspace{-0.02cm}\hat{\bm{\beta}}^{(\hspace{-0.01cm}k\hspace{-0.01cm})}\hspace{-0.02cm}\right\}\hspace{-0.03cm},
\end{equation}
where 
\begin{equation}
    \label{lastterm}
    \begin{split}
    &p(\mathbf{x}_{t+1},\mathbf{y}_t|\mathbf{x}_t,\bm{\beta}) = \mathbf{1}_{[\boldsymbol{a},\boldsymbol{b}]}(\bm{\xi}_t\hspace{-0.05cm}-\hspace{-0.05cm}\mathbf{h}_t(\bm{\zeta}_t))  \\
    &\hspace{0.3cm}\times \hspace{-0.05cm}
    \displaystyle\frac{\exp\hspace{-0.06cm}\left\{\hspace{-0.05cm}-\frac{1}{2}\hspace{-0.01cm}(\bm{\xi}_t\hspace{-0.08cm}-\hspace{-0.07cm}\mathbf{h}_t(\bm{\zeta}_t)\hspace{-0.06cm}-\hspace{-0.06cm}\bm{\mu})^{\hspace{-0.02cm}\top}\hspace{-0.02cm}\bm{\Sigma}^{\hspace{-0.01cm}-1}\hspace{-0.02cm}(\bm{\xi}_t\hspace{-0.08cm}-\hspace{-0.07cm}\mathbf{h}_t(\bm{\zeta}_t)\hspace{-0.08cm}-\hspace{-0.07cm}\bm{\mu})\hspace{-0.03cm}\right\}}{\int_{\boldsymbol{a}-\bm{\mu}}^{\boldsymbol{b}-\bm{\mu}} \exp\left\{-\frac{1}{2}\mathbf{p}^\top \bm{\Sigma}^{-1} \mathbf{p}\right\}\textnormal{d}\mathbf{p}}.    
    \end{split}
\end{equation}
    Lemma~\ref{lemma1} reformulates the optimization problem solved in each EM iteration to estimate the parameters $\bm{\mu}, \bm{\Sigma}, \boldsymbol{a}$, and $\boldsymbol{b}$.
\begin{lemma}
\label{lemma1}
    Define $\mathcal{A}_t:=\textnormal{supp}\big(\hspace{0.035cm}p(\mathbf{x}_t,\mathbf{x}_{t+1}|\mathbf{y}_{1:N},\hat{\bm{\beta}}^{(k)})\big)$. The next EM iteration of $\bm{\beta}$ given $\hat{\bm{\beta}}^{(k)}$ is given by the solution of the following optimization problem 
\begin{subequations}
\label{optimizationbeta}
        \begin{align}
        \label{optimizationbeta1}
        \hat{\bm{\beta}}^{(k+1)}=\hspace{0.1cm} &\underset{\bm{\beta}}{\arg \min} \hspace{0.08cm} \mathcal{V}^{(k)}(\bm{\beta}) \\
        \label{optimizationbeta2}
        &\hspace{-1.6cm}\textnormal{ s.t. }\hspace{-0.03cm}\boldsymbol{a}\hspace{-0.05cm}\leq\hspace{-0.03cm}\bm{\xi}_t\hspace{-0.05cm}-\hspace{-0.03cm}\mathbf{h}_t(\bm{\zeta}_t)\hspace{-0.03cm}\leq\hspace{-0.02cm} \boldsymbol{b} \hspace{-0.02cm}\textnormal{ for all }\hspace{-0.04cm} (\hspace{-0.01cm}\mathbf{x}_t,\hspace{-0.02cm}\mathbf{x}_{t\hspace{-0.01cm}+\hspace{-0.01cm}1}\hspace{-0.01cm})\hspace{-0.06cm}\in\hspace{-0.05cm}\mathcal{A}_t,  t\hspace{-0.06cm}=\hspace{-0.06cm}1,\hspace{-0.02cm}\dots,\hspace{-0.01cm}N\hspace{-0.02cm},
    \end{align}
\end{subequations}
where
\begin{subequations}
\label{vs}
    \begin{align}
    \label{v1}
    \hspace{-0.15cm}\mathcal{V}^{(k)}(\bm{\beta})\hspace{-0.04cm} &=\hspace{-0.04cm}\frac{1}{2}\textnormal{tr}\Big\{\bm{\Sigma}^{-1}\big(\hspace{0.03cm}\bm{\Phi}\hspace{-0.05cm}-\bm{\Psi}\bm{\mu}^\top-\bm{\mu}\bm{\Psi}^\top+\bm{\mu}\bm{\mu}^\top\big) \Big\} \\
    \label{v2}
    &\hspace{-0.5cm}+\log \hspace{-0.05cm}\int_{\boldsymbol{a}}^{\boldsymbol{b}} \hspace{-0.12cm}\exp\left\{\hspace{-0.03cm}-\frac{1}{2}(\mathbf{x}-\bm{\mu})^\top \bm{\Sigma}^{-1} \hspace{-0.03cm}(\mathbf{x}-\bm{\mu})\right\} \hspace{-0.04cm}\textnormal{d}\mathbf{x}, 
\end{align}
\end{subequations}
and where we have defined the following quantities:
\begin{subequations}
\label{smoothingmatrices}
\begin{align}
\label{psi}
\bm{\Psi} \hspace{-0.09cm}&=\hspace{-0.11cm} \frac{1}{N}\hspace{-0.08cm}\sum_{t=1}^N \hspace{-0.09cm}\int_{\hspace{-0.03cm}\mathbb{R}^{n}}\hspace{-0.1cm}\int_{\hspace{-0.03cm}\mathbb{R}^{n}} \hspace{-0.16cm}(\bm{\xi}_t\hspace{-0.09cm}-\hspace{-0.08cm}\mathbf{h}_t(\bm{\zeta}_t)) p(\mathbf{x}_{t\hspace{-0.01cm}+\hspace{-0.01cm}1}\hspace{-0.02cm},\hspace{-0.02cm}\mathbf{x}_t|\mathbf{y}_{\hspace{-0.01cm}1\hspace{-0.01cm}:\hspace{-0.01cm}N}\hspace{-0.02cm},\hspace{-0.02cm}\hat{\bm{\beta}}^{(\hspace{-0.01cm}k\hspace{-0.01cm})}\hspace{-0.02cm}) \textnormal{d}\mathbf{x}_{t\hspace{-0.01cm}+\hspace{-0.01cm}1}\textnormal{d}\mathbf{x}_t, \\
\bm{\Phi} \hspace{-0.06cm}&=\hspace{-0.05cm} \frac{1}{N}\sum_{t=1}^N \int_{\hspace{-0.03cm}\mathbb{R}^{n}}\hspace{-0.1cm}\int_{\hspace{-0.03cm}\mathbb{R}^{n}} (\bm{\xi}_t\hspace{-0.08cm}-\hspace{-0.07cm}\mathbf{h}_t(\bm{\zeta}_t))
(\bm{\xi}_t\hspace{-0.08cm}-\hspace{-0.07cm}\mathbf{h}_t(\bm{\zeta}_t))^\top \notag  \\ 
\label{phi}
&\hspace{0.8cm}\times p(\mathbf{x}_{t+1},\mathbf{x}_t|\mathbf{y}_{1:N},\hat{\bm{\beta}}^{(k)}) \textnormal{d}\mathbf{x}_{t+1}\textnormal{d}\mathbf{x}_t.
\end{align}
    
\end{subequations}
\end{lemma}
\vspace{0.15cm}
\textit{Proof}. After applying the conditional expectation with respect to $\boldsymbol{y}_{1:N}$ and $\hat{\bm{\beta}}^{(k)}$ to the log joint density in \eqref{lastterm}, we find that the maximizer of $\mathcal{Q}(\bm{\beta},\hat{\bm{\beta}}^{(k)})$ in \eqref{qfunction} must satisfy 
    \begin{equation}
        \mathbb{E}\{\log \mathbf{1}_{[\boldsymbol{a},\boldsymbol{b}]}(\bm{\xi}_t-\mathbf{h}_t(\bm{\zeta}_t))|\mathbf{y}_{1:N},\hat{\bm{\beta}}^{(k)}\}\geq C \notag 
    \end{equation}
    for some constant $C\in \mathbb{R}$ (otherwise, $\mathcal{Q}$ is ill-defined). Thus, we must have $\boldsymbol{a}\leq \bm{\xi}_t\hspace{-0.08cm}-\hspace{-0.07cm}\mathbf{h}_t(\bm{\zeta}_t) \leq \boldsymbol{b}$ for all $(\mathbf{x}_t,\mathbf{x}_{t+1})\in \textnormal{supp}\big(p(\mathbf{x}_t,\mathbf{x}_{t+1}|\mathbf{y}_{1:N},\hat{\bm{\beta}}^{(k)})\big)$. Under this constraint, we have
\begin{align}
    \mathbb{E}&\{\log \mathbf{1}_{[\boldsymbol{a},\boldsymbol{b}]}(\bm{\xi}_t\hspace{-0.04cm}-\hspace{-0.04cm}\mathbf{h}_t(\bm{\zeta}_t))|\mathbf{y}_{1:N},\hat{\bm{\beta}}^{(k)}\} \notag \\
    &\hspace{-0.31cm}= \hspace{-0.14cm}\int_{\hspace{-0.03cm}\mathbb{R}^{n}}\hspace{-0.1cm}\int_{\hspace{-0.03cm}\mathbb{R}^{n}}\hspace{-0.2cm} \log \hspace{-0.04cm}\mathbf{1}_{[\boldsymbol{a},\hspace{-0.02cm}\boldsymbol{b}]}(\bm{\xi}_t\hspace{-0.11cm}-\hspace{-0.08cm}\mathbf{h}_t(\bm{\zeta}_t)\hspace{-0.01cm}) p(\mathbf{x}_t,\hspace{-0.03cm}\mathbf{x}_{t\hspace{-0.01cm}+\hspace{-0.01cm}1}\hspace{-0.01cm}|\mathbf{y}_{1\hspace{-0.01cm}:N},\hspace{-0.02cm}\hat{\bm{\beta}}^{(k)}\hspace{-0.02cm}) \textnormal{d}\mathbf{x}_{t\hspace{-0.01cm}+\hspace{-0.01cm}1}\textnormal{d}\mathbf{x}_{t} \notag \\
    &\hspace{-0.31cm}= 0. \notag
\end{align}
On the other hand, after exploiting properties of the trace and the linearity of the expectation operator, 
\begin{align}
    &\sum_{t=1}^N \mathbb{E}\hspace{-0.04cm}\left\{\hspace{-0.04cm}(\bm{\xi}_t\hspace{-0.09cm}-\hspace{-0.06cm}\mathbf{h}_t(\bm{\zeta}_t)\hspace{-0.06cm}-\hspace{-0.06cm}\bm{\mu})^{\hspace{-0.04cm}\top}\hspace{-0.03cm}\bm{\Sigma}^{-\hspace{-0.02cm}1}\hspace{-0.04cm}(\bm{\xi}_t\hspace{-0.09cm}-\hspace{-0.06cm}\mathbf{h}_t(\bm{\zeta}_t)\hspace{-0.06cm}-\hspace{-0.06cm}\bm{\mu})\big|\mathbf{y}_{1:N},\hat{\bm{\beta}}^{(k)}\right\}\hspace{-0.07cm}= \notag \\
    & \textnormal{tr}\hspace{-0.075cm}\left\{ \hspace{-0.07cm}\bm{\Sigma}^{\hspace{-0.02cm}-\hspace{-0.02cm}1} \hspace{-0.07cm}\sum_{t=1}^N\hspace{-0.03cm}\mathbb{E}\hspace{-0.05cm}\left\{\hspace{-0.04cm}(\bm{\xi}_t\hspace{-0.09cm}-\hspace{-0.06cm}\mathbf{h}_t(\hspace{-0.01cm}\bm{\zeta}_t\hspace{-0.01cm})\hspace{-0.07cm}-\hspace{-0.07cm}\bm{\mu})\hspace{-0.02cm}(\bm{\xi}_t\hspace{-0.09cm}-\hspace{-0.065cm}\mathbf{h}_t(\hspace{-0.01cm}\bm{\zeta}_t\hspace{-0.01cm})\hspace{-0.07cm}-\hspace{-0.07cm}\bm{\mu})^{\hspace{-0.05cm}\top} \hspace{-0.02cm}\big|\mathbf{y}_{1\hspace{-0.01cm}:\hspace{-0.01cm}N}\hspace{-0.02cm},\hspace{-0.02cm}\hat{\bm{\beta}}^{(\hspace{-0.01cm}k\hspace{-0.01cm})}\hspace{-0.07cm}\right\} \hspace{-0.08cm}\right\}\hspace{-0.07cm}, \notag
\end{align}
which, after expanding the expectation, leads to \eqref{v1} with matrices given by \eqref{smoothingmatrices}. Since the integral term in the denominator in \eqref{lastterm} is deterministic and does not depend on $t$, we immediately obtain \eqref{v2}. \hfill $\square$

The constraints associated with the optimization problem in Lemma~\ref{lemma1} describe the parameter space where all state trajectories described by $\hat{\bm{\beta}}^{(k)}$ are feasible. The condition stated in \eqref{optimizationbeta2} is fulfilled when $\bm{\beta} = \hat{\bm{\beta}}^{(k)}$, indicating that $\mathcal{Q}(\bm{\beta},\hat{\bm{\beta}}^{(k)})$ is well defined in a region that contains $\hat{\bm{\beta}}^{(k)}$. As a result, there exists a parameter vector $\hat{\bm{\beta}}^{(k+1)}$ that satisfies these constraints and minimizes $\mathcal{V}^{(k)}(\bm{\beta})$.
\begin{remark}
    The optimization problem in Lemma \ref{lemma1} generalizes the unconstrained Gaussian case for linear time-invariant (LTI) systems in Lemma 3.1 of \cite{gibson2005robust}. Indeed, if $a_i\to -\infty$ and $b_i\to \infty$, the constraints in \eqref{optimizationbeta2} are always satisfied and the log-int-exp term in \eqref{v2} is replaced by a log-det expression by observing that, for any $\mathbf{S}\succ \mathbf{0}$,
    \begin{equation}
        \int_{\mathbb{R}^n} \exp\left\{-\frac{1}{2}\mathbf{x}^\top \mathbf{S}^{-1} \mathbf{x}\right\}\textnormal{d}\mathbf{x} = \sqrt{\det(2\pi \mathbf{S})}.
    \end{equation}
    Replacing accordingly and setting $\bm{\mu}=\mathbf{0}$ and $\boldsymbol{h}_t(\bm{\eta}_t)= \bm{\Gamma}\bm{\eta}_t$, i.e., the LTI case, we reach Eq. (22) of \cite{gibson2005robust}.
\end{remark}

Next, we characterize the solution of the optimization problem in \eqref{optimizationbeta}. The set of nonlinear equations that the parameters in $\bm{\beta}$ must satisfy at the optimal point are derived next, which constitutes Contribution~\ref{contributionC1} of this paper. To ease the notation, we have excluded the iteration indices on the parameters $\bm{\mu}, \bm{\Sigma}, \boldsymbol{a}$, and $\boldsymbol{b}$.
\begin{theorem}
\label{thmproblem1}
The global optimum of $\mathcal{Q}(\bm{\beta},\hat{\bm{\beta}}^{(k)})$ in \eqref{qfunction} with respect to $\bm{\beta}$ in \eqref{parameterbeta}, for fixed values of $\hat{\bm{\beta}}^{(k)}$ and $\mathbf{y}_{1:N}$, is given by $\hat{\bm{\beta}}^{(k+1)}=[\hat{\bm{\mu}}^\top, \textnormal{vec}\{\hat{\bm{\Sigma}}\}^\top, \hat{\boldsymbol{a}}^\top, \hat{\boldsymbol{b}}^\top]^\top$, where the parameters solve the following set of nonlinear equations:
\begin{subequations}\label{eq:system}
\begin{empheq}[left=\empheqlbrace]{align}
    \label{optimalm1}
    &\mathcal{M}_1(\hat{\bm{\mu}},\hat{\bm{\Sigma}}, \hat{\boldsymbol{a}}, \hat{\boldsymbol{b}})=\bm{\Psi}, \\
    \label{optimalm2}
    &\mathcal{M}_2(\hat{\bm{\mu}},\hat{\bm{\Sigma}}, \hat{\boldsymbol{a}}, \hat{\boldsymbol{b}})=\bm{\Phi}, \\
    \label{optimala}
     &\{\hat{\boldsymbol{a}}\}_i = \inf_{(\mathbf{x}_t,\mathbf{x}_{t+1})\in\mathcal{A}_t. t=1,\dots,N} \{\bm{\xi}_{t}-\mathbf{h}_t(\bm{\zeta}_{t})\}_i, \\
    \label{optimalb}
    &\{\hat{\boldsymbol{b}}\}_i = \sup_{(\mathbf{x}_t,\mathbf{x}_{t+1})\in\mathcal{A}_t. t=1,\dots,N} \{\bm{\xi}_{t}-\mathbf{h}_t(\bm{\zeta}_{t})\}_i,
\end{empheq}
\end{subequations}
where $\mathcal{M}_i(\bm{\mu},\bm{\Sigma}, \boldsymbol{a},\boldsymbol{b})$ represents the $i$th moment of a Gaussian with mean $\bm{\mu}$ and covariance $\bm{\Sigma}$ when truncated to a hyperrectangle with vertices $\boldsymbol{a}$ and $\boldsymbol{b}$, and where
$\bm{\Psi}$ and $\bm{\Phi}$ are given by \eqref{psi} and \eqref{phi}, respectively.
\end{theorem}
    \textit{Proof}. We begin by deriving the truncation bound updates \eqref{optimala} and \eqref{optimalb}. Since the optimal value $\hat{\bm{\beta}}^{(k+1)}$ must satisfy the constraints $\hat{\boldsymbol{a}}\leq \bm{\xi}_t\hspace{-0.08cm}-\hspace{-0.07cm}\mathbf{h}_t(\bm{\zeta}_t) \leq \hat{\boldsymbol{b}}$ for all $(\mathbf{x}_t,\mathbf{x}_{t+1})\in \textnormal{supp}\big(p(\mathbf{x}_t,\mathbf{x}_{t+1}|\mathbf{y}_{1:N},\hat{\bm{\beta}}^{(k)})\big)$, the following inequalities must hold for $i=1,\dots,n$:
\begin{align}
\label{aiinequality}
    \hspace{-0.3cm}\{\hspace{-0.01cm}\hat{\boldsymbol{a}}\hspace{-0.01cm}\}_i \hspace{-0.06cm}&\leq \hspace{-0.05cm}\inf\hspace{-0.06cm}\big\{\hspace{-0.04cm}\{\bm{\xi}_{t}\hspace{-0.07cm}-\hspace{-0.07cm}\mathbf{h}_t(\hspace{-0.01cm}\bm{\zeta}_{t}\hspace{-0.01cm})\hspace{-0.01cm}\}_i|(\mathbf{x}_t,\hspace{-0.02cm}\mathbf{x}_{t\hspace{-0.01cm}+\hspace{-0.01cm}1}\hspace{-0.01cm})\hspace{-0.09cm}\in\hspace{-0.08cm} \mathcal{A}_t,\hspace{-0.02cm}t\hspace{-0.07cm}=\hspace{-0.07cm}1,\dots\hspace{-0.03cm},\hspace{-0.03cm}N \hspace{-0.04cm}\big\}\hspace{-0.02cm},  \hspace{-0.1cm} \\
        \hspace{-0.3cm}\{\hat{\boldsymbol{b}}\}_i \hspace{-0.05cm}&\geq \hspace{-0.04cm}\sup\hspace{-0.06cm}\big\{\hspace{-0.03cm}\{\bm{\xi}_{t}\hspace{-0.06cm}-\hspace{-0.06cm}\mathbf{h}_t(\hspace{-0.01cm}\bm{\zeta}_{t}\hspace{-0.01cm})\hspace{-0.01cm}\}_i|(\mathbf{x}_t,\hspace{-0.02cm}\mathbf{x}_{t\hspace{-0.01cm}+\hspace{-0.01cm}1}\hspace{-0.01cm})\hspace{-0.06cm}\in\hspace{-0.06cm} \mathcal{A}_t,t\hspace{-0.06cm}=\hspace{-0.06cm}1,\dots\hspace{-0.02cm},\hspace{-0.03cm}N \hspace{-0.04cm}\big\}\hspace{-0.02cm}.\hspace{-0.1cm} \notag
\end{align}
We prove equality \eqref{optimala}; the same logic is applied to prove the result for $\hat{\boldsymbol{b}}$. To argue by contradiction, consider the global optimum $\hat{\bm{\beta}}^{(k+1)}$ and assume that there exists some $i^*\in\{1,\dots,n\}$ such that $\{\hat{\boldsymbol{a}}\}_{i^*}$ is strictly less than the right-hand side of \eqref{aiinequality}. Let $\tilde{\bm{\beta}}^{(k+1)}$ (with lower truncation limit $\tilde{\boldsymbol{a}}$) be the parameter vector that is equal to $\hat{\bm{\beta}}^{(k+1)}$ in all entries except $\{\hat{\boldsymbol{a}}\}_{i^*}$, at which it is equal to the right-hand side of \eqref{aiinequality}. This parameter vector also satisfies the constraints of the optimization problem in \eqref{optimizationbeta2}. Since any Gaussian PDF has support on $\mathbb{R}^{n}$, we have the strict inequality
\begin{equation}
    \int_{\hat{\boldsymbol{a}}-\bm{\mu}}^{\hat{\boldsymbol{b}}-\bm{\mu}} \hspace{-0.05cm}\exp\hspace{-0.03cm}\left\{\hspace{-0.04cm}-\hspace{-0.01cm}\frac{1}{2}\hspace{-0.01cm}\mathbf{x}^{\hspace{-0.02cm}\top}\hspace{-0.02cm} \bm{\Sigma}^{-\hspace{-0.02cm}1} \hspace{-0.02cm}\mathbf{x}\hspace{-0.05cm}\right\} \hspace{-0.03cm}\textnormal{d}\mathbf{x}\hspace{-0.03cm}>\hspace{-0.05cm}\int_{\tilde{\boldsymbol{a}}-\bm{\mu}}^{\hat{\boldsymbol{b}}-\bm{\mu}} \hspace{-0.12cm}\exp\hspace{-0.05cm}\left\{\hspace{-0.05cm}-\hspace{-0.01cm}\frac{1}{2}\hspace{-0.01cm}\mathbf{x}^{\hspace{-0.02cm}\top}\hspace{-0.04cm} \bm{\Sigma}^{-\hspace{-0.02cm}1} \hspace{-0.02cm}\mathbf{x}\hspace{-0.05cm}\right\} \hspace{-0.04cm}\textnormal{d}\mathbf{x}\hspace{-0.01cm}, \notag
\end{equation}
which leads to $\mathcal{V}^{(k)}(\hat{\bm{\beta}}^{(k+1)})>\mathcal{V}^{(k)}(\tilde{\bm{\beta}}^{(k+1)})$, i.e., $\hat{\bm{\beta}}^{(k+1)}$ is not the global optimum. This proves \eqref{optimala}.

Fixing $\boldsymbol{a}=\hat{\boldsymbol{a}}$ and $\boldsymbol{b}=\hat{\boldsymbol{b}}$, the optimizer is characterized by the first-order optimality conditions of the unconstrained problem in \eqref{optimizationbeta}. The following gradients are derived using standard matrix differentiation rules (see, e.g., \cite{magnus2019matrix}):
\begin{align}
\frac{\partial \mathcal{V}^{(k)}\hspace{-0.03cm}(\bm{\beta})}{\partial \bm{\mu}} \hspace{-0.07cm}&=\hspace{-0.05cm}\bm{\Sigma}^{-1}\hspace{-0.13cm}\left[\hspace{-0.03cm}\bm{\Psi}\hspace{-0.06cm}-\hspace{-0.06cm}\bm{\mu}\hspace{-0.06cm}-\hspace{-0.06cm}\frac{ \int_{\hat{\boldsymbol{a}}-\bm{\mu}}^{\hat{\boldsymbol{b}}-\bm{\mu}}\mathbf{x}\hspace{-0.02cm}\exp\{-\frac{1}{2}\mathbf{x}^\top \bm{\Sigma}^{-1}\mathbf{x}\}\textnormal{d}\mathbf{x}}{\int_{\hat{\boldsymbol{a}}-\bm{\mu}}^{\hat{\boldsymbol{b}}-\bm{\mu}}\exp\{-\frac{1}{2}\mathbf{x}^\top \bm{\Sigma}^{-1}\mathbf{x}\}\textnormal{d}\mathbf{x}}\right]\hspace{-0.06cm}, \notag \\
    \frac{\partial \mathcal{V}^{(k)}\hspace{-0.02cm}(\bm{\beta})}{\partial \bm{\Sigma}} \hspace{-0.07cm}&=\hspace{-0.07cm} -\frac{1}{2}\bm{\Sigma}^{-1}\big( \hspace{0.03cm}\bm{\Phi}\hspace{-0.05cm}-\bm{\Psi}\bm{\mu}^\top-\bm{\mu}\bm{\Psi}^\top+\bm{\mu}\bm{\mu}^\top \big)\bm{\Sigma}^{-1} \notag \\
    \label{replacing}
&\hspace{-0.4cm}+\hspace{-0.05cm}\frac{1}{2}\bm{\Sigma}^{-1}\frac{ \int_{\hat{\boldsymbol{a}}-\bm{\mu}}^{\hat{\boldsymbol{b}}-\bm{\mu}}\mathbf{x}\mathbf{x}^{\top}\hspace{-0.06cm}\exp\{-\frac{1}{2}\mathbf{x}^\top \bm{\Sigma}^{-1}\mathbf{x}\}\textnormal{d}\mathbf{x}}{\int_{\hat{\boldsymbol{a}}-\bm{\mu}}^{\hat{\boldsymbol{b}}-\bm{\mu}}\exp\{-\frac{1}{2}\mathbf{x}^\top \bm{\Sigma}^{-1}\mathbf{x}\}\textnormal{d}\mathbf{x}}\bm{\Sigma}^{-1}. \hspace{-0.1cm}
\end{align}
Note that the integral expressions above can be written in terms of the moments of a truncated Gaussian distribution as follows:
\begin{align}
    \frac{ \int_{\hat{\boldsymbol{a}}\hspace{-0.02cm}-\hspace{-0.02cm}\bm{\mu}}^{\hat{\boldsymbol{b}}\hspace{-0.02cm}-\hspace{-0.02cm}\bm{\mu}}\mathbf{x}\hspace{-0.02cm}\exp\{\hspace{-0.02cm}-\frac{1}{2}\mathbf{x}^{\hspace{-0.03cm}\top} \bm{\Sigma}^{-1}\mathbf{x}\}\textnormal{d}\mathbf{x}}{\int_{\hat{\boldsymbol{a}}-\bm{\mu}}^{\hat{\boldsymbol{b}}-\bm{\mu}}\exp\{-\frac{1}{2}\mathbf{x}^\top \bm{\Sigma}^{-1}\mathbf{x}\}\textnormal{d}\mathbf{x}} \hspace{-0.06cm}&= \mathcal{M}_1- \bm{\mu}, \notag \\
    \frac{ \int_{\hat{\boldsymbol{a}}\hspace{-0.02cm}-\hspace{-0.02cm}\bm{\mu}}^{\hat{\boldsymbol{b}}\hspace{-0.02cm}-\hspace{-0.02cm}\bm{\mu}}\hspace{-0.05cm}\mathbf{x}\mathbf{x}^{\hspace{-0.04cm}\top}\hspace{-0.09cm} \exp\{\hspace{-0.03cm}-\hspace{-0.02cm}\frac{1}{2}\mathbf{x}^{\hspace{-0.045cm}\top}\hspace{-0.025cm} \bm{\Sigma}^{-\hspace{-0.025cm}1}\hspace{-0.01cm}\mathbf{x}\}\textnormal{d}\mathbf{x}}{\int_{\hat{\boldsymbol{a}}-\bm{\mu}}^{\hat{\boldsymbol{b}}-\bm{\mu}}\exp\{-\frac{1}{2}\mathbf{x}^\top \bm{\Sigma}^{-1}\mathbf{x}\}\textnormal{d}\mathbf{x}} \hspace{-0.075cm}&= \hspace{-0.075cm}\mathcal{M}_{\hspace{-0.01cm}2}\hspace{-0.08cm}-\hspace{-0.09cm} \mathcal{M}_{\hspace{-0.02cm}1}\bm{\mu}^{\hspace{-0.04cm}\top}\hspace{-0.155cm}-\hspace{-0.095cm}\bm{\mu}^{\hspace{-0.04cm}\top}\hspace{-0.08cm} \mathcal{M}_{\hspace{-0.02cm}1}\hspace{-0.085cm} +\hspace{-0.07cm} \bm{\mu}\bm{\mu}^{\hspace{-0.045cm}\top}\hspace{-0.08cm}, \notag     
\end{align}
where $\mathcal{M}_1$ and $\mathcal{M}_2$ are the shorthand notation for $\mathcal{M}_1(\bm{\mu},\bm{\Sigma}, \hat{\boldsymbol{a}}, \hat{\boldsymbol{b}})$ and $\mathcal{M}_2(\bm{\mu},\bm{\Sigma}, \hat{\boldsymbol{a}}, \hat{\boldsymbol{b}})$, respectively. Setting the gradient with respect to $\bm{\mu}$ to zero, we directly obtain \eqref{optimalm1}. When this result is inserted in \eqref{replacing}, with the gradient with respect to $\bm{\Sigma}$ being set to zero, we reach \eqref{optimalm2}. This concludes the proof. \hfill $\square$

Theorem \ref{thmproblem1} describes the iterations of the proposed EM algorithm. For a fixed iteration $\hat{\bm{\beta}}^{(k)}$, Eqs. \eqref{optimalm1} and \eqref{optimalm2} can be interpreted as method of moments equations \cite{newey1994large}, while \eqref{optimala} and \eqref{optimalb} describe the smallest support for $\bm{\eta}_t$ that can explain the dataset under the assumption that the noise model is described exactly by $\hat{\bm{\beta}}^{(k)}$.

\section{Implementation Aspects}
\label{sec:implementation}
Section \ref{sec:truncated} outlined the theoretical foundations of the proposed identification method; here, important implementation aspects of the approach are discussed.

\subsection{Initialization}
A good initial estimate $\hat{\bm{\beta}}_1$ can significantly reduce the iterations required for EM algorithm to converge, and may avoid local optima. In many cases, an adequate initialization can be computed by solving the set of equations of Theorem \ref{thmproblem1} for the untruncated Gaussian case, i.e., $a_i\to -\infty$ and $b_i\to \infty$ for $i=1,\dots,n\hspace{-0.03cm}+\hspace{-0.03cm}p$. This yields $\hat{\mathcal{M}}_1 \hspace{-0.03cm}=\hspace{-0.03cm} \mathbf{0}$ and $\hat{\mathcal{M}}_2 \hspace{-0.03cm}=\hspace{-0.03cm} \hat{\bm{\Sigma}}-\hat{\bm{\mu}}\hat{\bm{\mu}}^\top$, leading to the initial estimate
\begin{equation}
\hat{\bm{\mu}}_1 = \bm{\Psi}, \quad \hat{\bm{\Sigma}}_1 = \bm{\Phi}-\bm{\Psi}\bm{\Psi}^\top. \notag
\end{equation}
Thus, initializing the proposed EM approach with estimates from the Kalman smoothing method in \cite{gibson2005robust} can lead to a fast convergence of the iterations in Theorem \ref{thmproblem1}. This strategy is preferred  when the Gaussian distribution being truncated has most of its probability density concentrated within the hyperrectangle $[\boldsymbol{a},\boldsymbol{b}]$.

\subsection{Computation of sample and population moments}
Equations \eqref{optimalm1} and \eqref{optimalm2} require the computation of sample and population moments. For the sample moments $\bm{\Psi}$ and $\bm{\Phi}$, a cornerstone of the proposed method is the smoothing procedure that must be performed. Sequential Monte Carlo methods \cite{doucet2000sequential}, also known as particle filtering and smoothing, provide a framework for estimating the state of a dynamic system based on noisy observations. The key idea is to represent the posterior distribution of the system state using a set of particles and weights, that is 
\begin{align}
    p(\tilde{\mathbf{x}}_{t}|\mathbf{y}_{1:N}) &\approx \sum_{i=1}^{M} w_{t|N}^{(i)} \delta\hspace{-0.02cm}\big( \tilde{\mathbf{x}}_t-\tilde{\mathbf{x}}_t^{(i)}\big) , \notag 
\end{align}
where $\delta\left(\cdot\right)$ denotes the Dirac delta function, $\tilde{\mathbf{x}}_{t}:=[\mathbf{x}_{t+1}^\top,\mathbf{x}_{t}^\top]^\top$ is the augmented state vector, $w_{t|N}^{(i)}$ and $\tilde{\mathbf{x}}_t^{(i)}$ are the $i$th weight and particle that represent the smoothing distribution $p(\tilde{\mathbf{x}}_{t}|\mathbf{y}_{1:N})$, and $M$ is the number of particles. Using the particle-based representation of the posterior PDF, any conditional expectation depending on the augmented state can be computed using 
\begin{equation}\label{eqn:gexpected}
    \mathbb{E}\left\lbrace \boldsymbol{f}(\tilde{\mathbf{x}}_t)|\mathbf{y}_{1:N}\right\rbrace \approx \sum_{i=1}^{M}w_{t|N}^{(i)}\mathit{f}(\tilde{\mathbf{x}}_t^{(i)}),
\end{equation}
where $\boldsymbol{f}(\tilde{\mathbf{x}}_t)$ is a function of the augmented state $\tilde{\mathbf{x}}_t$. In our setting, $\bm{\Psi}$ and $\bm{\Phi}$ are computed using $\boldsymbol{f}(\tilde{\mathbf{x}}_t)= \bm{\xi}_t-\mathbf{h}_t(\bm{\zeta}_t)$ and $\boldsymbol{f}(\tilde{\mathbf{x}}_t)= (\bm{\xi}_t-\mathbf{h}_t(\bm{\zeta}_t))(\bm{\xi}_t-\mathbf{h}_t(\bm{\zeta}_t))^\top$, respectively.

With regards to the population moments, Theorem~\ref{thmproblem1} requires the computation of the first and second moments of a truncated Gaussian distribution. For any dimension, these can be computed using Monte Carlo integration and specialized tools for sampling high-dimensional truncated Gaussians \cite{botev2017normal}. For the univariate case, certain explicit formulas can be obtained; non-recursive expressions (in terms of the Gamma function) can be found in \cite{ogasawara2022unified}. For simplicity, we adopt the moment-generating function approach of \cite{tallis1961moment}.

For $k \in \mathbb{N}_0$, consider the computation of
	\begin{equation}
		\label{xk}
		\mathcal{M}_k\hspace{-0.01cm}(\mu,\hspace{-0.01cm}\sigma^2\hspace{-0.02cm},\hspace{-0.01cm} a,\hspace{-0.01cm} b) \hspace{-0.07cm}= \hspace{-0.09cm}\frac{1}{\sqrt{\hspace{-0.01cm}2\pi \hspace{-0.01cm}\sigma^2}} \hspace{-0.06cm}\int_a^b \hspace{-0.1cm}x^k \hspace{-0.04cm}\exp\hspace{-0.04cm}\left\lbrace \hspace{-0.03cm}-\hspace{-0.01cm}\frac{(x\hspace{-0.02cm}-\hspace{-0.02cm}\mu)^2}{2\sigma^2}\hspace{-0.03cm}\right\rbrace \hspace{-0.02cm} \textnormal{d}x.
	\end{equation}
	We introduce the moment generating function
	\begin{equation}
		M(t) := \frac{1}{\sqrt{2\pi \sigma^2}} \int_a^b \exp\left\lbrace tx\right\rbrace \exp\left\lbrace -\frac{(x-\mu)^2}{2\sigma^2}\right\rbrace \textnormal{d}x. \notag 
	\end{equation}
	For $t=0$ (i.e., $k=0$ in \eqref{xk}), it is known that $\mathcal{M}_0(\mu,\sigma^2, a, b) = M(0) = \left[\textnormal{erf}\left(\bar{b}\right)-\textnormal{erf}\left(\bar{a}\right)\right]/2$, where $\bar{b}=(b-\mu)/\sqrt{2\sigma^2}$ and $\bar{a}=(a-\mu)/\sqrt{2\sigma^2}$, and
	\begin{equation}
	\textnormal{erf}(x) = \frac{2}{\sqrt{\pi}} \int_0^x \exp\left\lbrace -z^2 \right\rbrace \textnormal{d}z. \notag
	\end{equation} 
	Moreover, $\mathcal{M}_k(\mu,\sigma^2, a, b) = M^{(k)}(0)$. By completing the square in the exponent of the integrand of $M(t)$,
	\begin{align}
		M(t) &=\frac{\varrho(t)}{\sqrt{2\pi \sigma^2}} \int_a^b \exp\left\lbrace -\frac{(x-[\mu+t\sigma^2])^2}{2\sigma^2}\right\rbrace \textnormal{d}x \notag \\
		&= \dfrac{\varrho(t)}{2}\left[\textnormal{erf}\left(\bar{b}-\frac{t\sigma}{\sqrt{2}}\right)-\textnormal{erf}\left(\bar{a}-\frac{t\sigma}{\sqrt{2}}\right)\right], \notag 
	\end{align}
	where $\varrho(t)=\exp\left\lbrace (t^2 \sigma^2/2)+\mu t \right\rbrace$. Since $\frac{\textnormal{d}}{\textnormal{d}t} \textnormal{erf}(\bar{\mu}+ct) =\frac{2c}{\sqrt{\pi}}\exp\{-(\bar{\mu}+ct)^2\}$, we see that $M^{(k)}(0)$ can be explicitly computed in terms of the exponential and erf functions. In particular, this leads to the explicit computation of the first and second moments
	\begin{align}
		\mathcal{M}_1(\mu,\sigma^2, a, b)&= \mu \mathcal{M}_0(\mu,\sigma^2, a, b)\notag \\
  &-\sigma \left(\exp\left\lbrace -\bar{b}^2\right\rbrace- \exp\left\lbrace -\bar{a}^2\right\rbrace \right)/\sqrt{2\pi}, \notag \\
	    \mathcal{M}_2(\mu,\sigma^2, a, b)&= (\mu^2+\sigma^2) \mathcal{M}_0(\mu,\sigma^2, a, b)\notag \\
     &\hspace{-2cm}-\sigma \left[(\mu+b)\exp\left\lbrace - \bar{b}^2\right\rbrace -(\mu+a)\exp\left\lbrace -\bar{a}^2\right\rbrace \right]/\sqrt{2\pi}. \notag
	\end{align}

\subsection{Computational methods for solving Eqs. in (15)}
Once the smoothing step has been performed, upper and lower bounds for \eqref{optimala} and \eqref{optimalb} can respectively be obtained by evaluating $\bm{\xi}_{t}-\mathbf{h}_t(\bm{\zeta}_{t})$ at the particle values $\tilde{\mathbf{x}}_t^{(i)}$, and computing the maximum and minimum of each entry. Caution must be exercised, as such computations scale with the number of particles and data points.

Once $\hat{\boldsymbol{a}}$ and $\hat{\boldsymbol{b}}$ are determined, or if the truncation limits are known, the remaining task is to solve the system in \eqref{optimalm1}-\eqref{optimalm2}. Following the approach in \cite{lee2012algorithms} for fitting truncated Gaussian mixtures, we employ the fixed-point iterations:
\begin{align}
    \hat{\bm{\mu}}_{j+1}&=\bm{\Psi}-\mathcal{M}_1(\mathbf{0},\hat{\bm{\Sigma}}_j, \hat{\boldsymbol{a}}-\hat{\bm{\mu}}_{j}, \hat{\boldsymbol{b}}-\hat{\bm{\mu}}_{j}), \notag \\
    \hat{\bm{\Sigma}}_{j+1}&= \bm{\Phi} + \hat{\bm{\Sigma}}_{j} - \mathcal{M}_2(\mathbf{0},\hat{\bm{\Sigma}}_j, \hat{\boldsymbol{a}}-\hat{\bm{\mu}}_{j+1}, \hat{\boldsymbol{b}}-\hat{\bm{\mu}}_{j+1}),  \notag 
\end{align}
where $\hat{\bm{\mu}}_{j}$ and $\hat{\bm{\Sigma}}_{j}$ represent the $j$th iterates of the mean and covariance within the $k$th EM iteration.

\section{Simulation example}
\label{sec:simulations}
In this section, we illustrate the proposed approach via a simulation example. We study the gain in estimation accuracy when estimating truncated Gaussian noise PDFs instead of standard Gaussian ones on the system
\begin{align}
    x_{t+1}&= f(x_t,u_t) + w_t, \notag \\
    y_t &=  g(x_t,u_t) + v_t, \notag
\end{align}
where $f(x_t,u_t)=0.9x_t + 2u_t$, $g(x_t,u_t)=1.6x_t + 1.2 u_t$, the noises $w_t\sim \mathcal{TN}(-0.3,1,-1.5,2.5)$ and $v_t\sim \mathcal{N}(-0.1,0.5)$ are uncorrelated. For simplicity we assume that the truncation bounds related to $w_t$ are known, and we use this information to estimate the mean and covariance of the Gaussian being truncated to generate $w_t$ (i.e., $-0.3$ and $1$ respectively), as well as the mean and covariance of $v_t$. 

A Monte Carlo simulation is performed with $100$ runs, each with $N=5000$ samples. Two estimators are tested: the EM algorithm that uses the Kalman smoother for computing the noise statistics (KS-EM), which is based on the work of \cite{gibson2005robust}, and the proposed method (TG-EM). A total of $40$ EM iterations is performed for each method, and they are initialized at a uniformly-distributed random value centered at the true parameters with $10\%$ relative error. Each smoothed PDF is approximated using $500$ particles. 

\begin{figure}
	\centering{
		\includegraphics[width=0.47\textwidth]{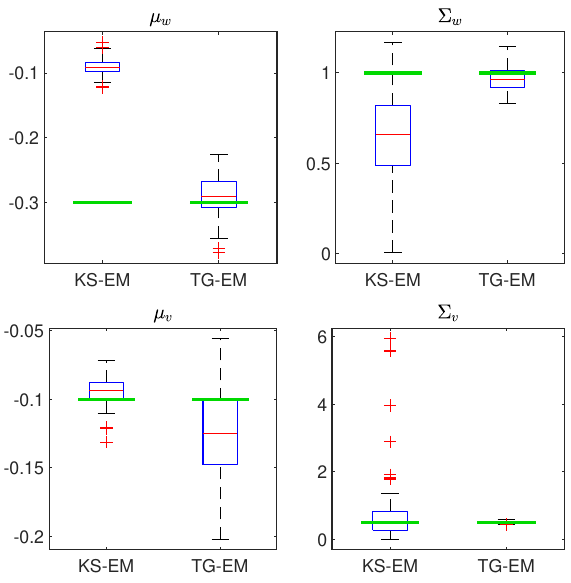}
    \caption{Boxplots of each estimated mean and variance of the state and output noise. The green line in each boxplot represents the true value of the parameter being estimated.}
    \vspace{-0.3cm}
		\label{fig_boxplot}}
\end{figure}

Boxplots containing the statistical performance of the KS-EM and TG-EM methods for each parameter of interest are presented in Figure~\ref{fig_boxplot}. The proposed method shows clear improvements in estimation accuracy over the Kalman-based approach in 3 of the 4 parameters. Interestingly, the median of the KS-EM estimates of $\mu_{w}$ and $\Sigma_w$ are approximately equal to the mean and variance of $w_t$, which are given by $-0.09$ and $0.66$, respectively. This is not the case for the proposed method, since the parameters $\mu_{w}$ and $\Sigma_w$ describe the mean and covariance of the Gaussian distribution prior to truncation, and not the distribution measures of the truncated PDF itself. Only the proposed approach can adequately incorporate the truncation bounds into the estimation, leading to a more accurate characterization of the noise in the state-space model.
The KS-EM estimates of $\mu_{v}$ are more accurate than those of the TG-EM method, however, the KS-EM estimates of $\Sigma_v$ show more spread and contain more outliers.

\section{Conclusions}
\label{sec:conclusions}
In this paper, we proposed a method for computing the parameters of a truncated Gaussian noise model that affects the state and output equations of a state-space system. This approach offers flexibility in noise model estimation, as combining Gaussianity with bounded support assumptions typically provides a more accurate characterization of the noise. By considering the state vector as a hidden variable in the EM algorithm, we derived explicit expressions that the parameters must satisfy at each iteration. This method generalizes the approach in \cite{gibson2005robust} for fixed system parameters, extending it from untruncated to truncated Gaussian distributions, with the option to set the truncation bounds to infinity, thus recovering the untruncated case. The computation of the EM iterations requires particle smoothing techniques, as well as the estimation of moments of truncated Gaussian PDFs. Future work includes accelerating the estimation procedure, and extending it to joint system and noise identification.

\bibliography{references}                 
\end{document}